\documentclass{Rinton-P9x6}

\begin{document}

\def\ra{\rangle}
\def\la{\langle}
\def\bege{\begin{equation}}
\def\ende{\end{equation}}
\def\begarr{\begin{eqnarray}}
\def\endarr{\end{eqnarray}}
\def\ha{{\hat a}}
\def\hb{{\hat b}}
\def\hu{{\hat u}}
\def\hv{{\hat v}}
\def\hc{{\hat c}}
\def\hd{{\hat d}}
\def\no{\noindent}
\def\non{\nonumber}
\def\hi{\hangindent=45pt}
\def\v{\vskip 12pt}
\def\unity{\mbox{\small 1} \!\! \mbox{1}}

\title{Quantum Imaging and Metrology}

\author{Hwang Lee, Pieter Kok, and Jonathan P. Dowling}

\address{
Exploration Systems Autonomy, Section 367, MS 126-347 \\
Quantum Computing Technologies Group \\
 Jet Propulsion Laboratory, 
 California Institute of Technology \\
 4800 Oak Grove Drive, Pasadena, CA~~91109-8099
}


\maketitle

\abstracts{
 The manipulation of quantum entanglement has found enormous potential
 for improving performances of devices such as gyroscopes, clocks, and
 even computers. Similar improvements have been demonstrated for
 lithography and microscopy. We present an overview of some aspects of
 enhancement by quantum entanglement in imaging and metrology.
}


In state-of-the-art optical-lithographic, semiconductor etching
techniques, the Rayleigh diffraction limit puts a lower bound on the
feature size that can be printed on a chip. This limit states that the
minimal resolvable feature is on the order of $\lambda/4$, where
$\lambda$ is the wavelength of the light used. Classically, if you
want to etch features of size 50~nm and smaller, you will be forced
to use optical radiation with wavelengths less than 200~nm. Hence, in
the optical lithographic community great efforts are put into producing 
commercial lithographic schemes that utilize wavelengths in the hard
UV and X-ray regimes\cite{yablo99}. However, such an approach
introduces severe technological and commercial difficulties. For
example, mirrors and lenses that are cheap and well understood in the
optical regime are much less common and much harder (and more
expensive) to make in the UV and X-ray regions of the
spectrum, and the problems become worse the shorter you go. Recently,
it has been shown, however, that the Rayleigh diffraction limit in
optical lithography can be circumvented by the use of path-entangled
photon number states\cite{boto00}.  

Fundamentally, optical light beams used for lithography are
quantum-mechanical in nature: they are superpositions of photon-number
states. This quantum language allows us to taylor non-classical states
of light, in which it is possible to program nonlocal correlations
between photons. Typically, one photon contains information about the
location and the momentum of other photons in the stream\cite{shih01}. 
If the proper nonlocal correlations are employed, you can actually
manipulate the location at which the light strikes the lithographic
substrate, such that features of size $\lambda/4N$ can be etched using
$N$ photons of wavelength $\lambda$. The use of such an effect was
also proposed in sub-natural spectroscopy\cite{rathe95}.

This remarkable property of quantum-correlated photons has been
recognized for some years in the context of quantum optical
interferometers. In a typical optical interferometer in which ordinary
coherent) laser light enters via only one port, the phase sensitivity
in the shot noise limit scales as $\Delta \varphi = 1/\sqrt{\bar n}$
where ${\bar n}$ is the mean number of photons\cite{scully97}. It
would seem that any desired sensitivity $\Delta \varphi$ could be
attained by simply increasing the laser power. However, when the
intensity of the laser (${\bar n}$) becomes too large, the power
fluctuations at the interferometer's mirrors introduce additional
noise terms that limit the device's overall sensitivity. 
Much of the early interest in squeezed
light empathized overcoming this signal-to-noise barrier. In the early
1980s it was demonstrated that squeezing  the vacuum in the unused
input port of the interferometer causes the  phase sensitivity to beat
the standard shot-noise limit\cite{caves81}. The total laser power
required for a given amount of phase sensitivity $\varphi$ is thus
greatly reduced. 

In 1986, Yurke and collaborators, as well as Yuen, considered the
question of phase noise reduction using correlated particles in number
states, incident upon both input ports of a Mach-Zehnder
interferometer\cite{yurke86,yurke86b,yuen86}. They showed that if $N$
quantum particles entered into each input port of the interferometer
in nearly equal numbers (and in a highly entangled fashion), then, for
large $N$, it was indeed possible to obtain an asymptotic phase
sensitivity of $1/N$, instead of $1/\sqrt{N}$. This is the best you
can do using number states in only one input port, and it indicates
that the photon counting noise does not originate from the intensity
fluctuations of the input beam\cite{scully93}. Similar observations
were made by many authors for optical
interferometers\cite{holland93,hillery93,brif96,kim98} as well as
Ramsey-type atom interferometers
\cite{yamamoto95,bollinger96,bouyer97,dowling98}. Wineland and
co-workers, in particular, have shown that the optimal frequency
measurement can be achieved by using {\it maximally entangled
  states}\cite{bollinger96}. These maximally entangled states are of a
particular interest, since they have a similar form as the ones
required for quantum lithography.

Let us take a look at the quantum enhancement due to maximally
entangled states, using standard parameter estimation. Consider an
ensemble of $N$ two-state systems in the state:
\be
|\varphi\rangle = {1 \over \sqrt{2}}
(|0\rangle+e^{i\varphi}|1\rangle)
\ee
\no
where $|0\rangle$ and $|1\rangle$ denote the two basis states. The
phase information can be obtained by measurement of an observable
$\hat{A} =|0\rangle\langle 1| + |1\rangle\langle 0|$. The expectation
value of $\hat{A}$ is then given by
\begin{equation}
  \langle\varphi|\hat{A}|\varphi\rangle=\cos\varphi\; .
\end{equation}
\no
When we repeat this experiment $N$ times, we obtain
 $ \langle\varphi_R|\hat{A}_R|\varphi_R\rangle=N\cos\varphi,
$
where
$
|\varphi_R\rangle = |\varphi\rangle_1 \ldots |\varphi\rangle_N
$,
and 
$\hat{A}_R = {\mbox{\Large $\oplus$}}_{k=1}^N 
  \hat{A}^{(k)}
$.
Since $\hat{A}_R^2=\unity$, the variance of $\hat{A}_R$, given $N$
samples, is readily computed to be $(\Delta A_R)^2 =
N(1-\cos^2\varphi) = N \sin^2 \varphi$. According to estimation
theory\cite{helstrom76}, we have  
\begin{equation}\label{est}
  \Delta\varphi_{\rm SL} = \frac{\Delta A_R}{|d\langle
  \hat{A}_R\rangle/d\varphi|} = \frac{1}{\sqrt{N}}\; .
\end{equation}
This is the standard variance in the parameter $\varphi$ after $N$
trials. In other words, the uncertainty in the phase is inversely
proportional to the square root of the number of trials. 
This is called the {\em shot-noise limit}.

Now consider an entangled state
\begin{equation}\label{entang}
 |\varphi_N\rangle\equiv {1 \over \sqrt{2}}
 |N,0\rangle + e^{iN\varphi}|0,N\rangle\; , 
\end{equation}
where $|N,0\rangle$ and $|0,N\rangle$ are 
collective states of $N$ particles, defined as
\begarr
 |N,0\ra &=& |0\ra_1 |0\ra_2 \cdot\cdot\cdot |0\ra_N \non \\
 |0,N\ra &=& |1\ra_1 |1\ra_2 \cdot\cdot\cdot |1\ra_N .
 \label{noon}
\endarr
The relative phase $e^{iN\varphi}$ is accumulated when each particle
in state $|1\ra$ acquires  a phase shift of $e^{i \varphi}$. An
important question now is what we need to measure in order to extract
the phase information. Recalling the single-particle case of ${\hat
  A}=|0\ra\la 1| + |1\ra \la 0|$, we need an observable that does what
the operator $|0,N\rangle\langle N,0| + |N,0\rangle\langle 0,N|$
does. For the given state of Eq.~(\ref{entang}), we can see that this
can be achieved by an observable, $\hat{A}_N = {\mbox{\Large
$\otimes$}}_{k=1}^N \hat{A}^{(k)} $. The expectation value of
$\hat{A}_N$ is then 
\begin{equation}\label{cosn}
  \langle\varphi_N |\hat{A}_N| \varphi_N\rangle = \cos N\varphi\; .  
\end{equation}
Again, $\hat{A}_N^2=\unity$, and $(\Delta A_N)^2 = 1-\cos^2 N\varphi =
\sin^2 N\varphi$. Using Eq.\ (\ref{est}) again, we obtain the
so-called Heisenberg limit (HL) of the minimal detectable phase:  
\begin{equation}\label{bol}
  \Delta\varphi_{\rm HL} = \frac{\Delta A_N}{|d\langle \hat{A}_N
  \rangle/d\varphi|}=\frac{1}{N}\; .
\end{equation}
The precision in $\varphi$ is increased by a factor $\sqrt{N}$ over
the standard noise limit. Of course, the preparation of a quantum
state such as Eq.~(\ref{entang}) is essential to the given
protocol\cite{lee02b}. 

In quantum lithography, we exploit the $\cos N\varphi$ behavior, 
exhibited by Eq.~(\ref{cosn}), to draw closely spaced lines on a
suitable substrate\cite{boto00}. Entanglement-enhanced frequency
measurements\cite{bollinger96} and gyroscopy\cite{dowling98} exploit
the $\sqrt{N}$ increased precision given by Eq.~(\ref{bol}). 
The physical interpretations of $A_N$ and the phase $\varphi$ will
differ in the different protocols. Three distinct physical
representations of this construction are of particular
interest.
 
First, in a Mach-Zehnder interferometer [as depicted in Fig.~1(a)] the
input light field is divided into two different paths by a beam
splitter, and recombined by another beam splitter. The phase
difference between the two paths is then measured by balanced
detection of the two output modes. Secondly, in Ramsey-type
spectroscopy, atoms are put in a superposition of the  ground state
and an excited state with a $\pi/2$-pulse [see Fig.~1(b)].  After a
relative phase shift is accumulated by the atomic states during the free
evolution, the second $\pi/2$-pulse is applied and the internal state
of the outgoing atom is measured. The third system is given by a qubit
that undergoes a Hadamard transform $H$, then picks up a relative
phase and is transformed back with a second Hadamard transformation
[Fig.~1(c)]. This representation is more mathematical than the
previous two, and it allows us to extract the unifying mathematical
principle that underlies the three systems. 

\begin{figure}[t]
\epsfxsize=20pc 
\begin{center}
\epsfbox{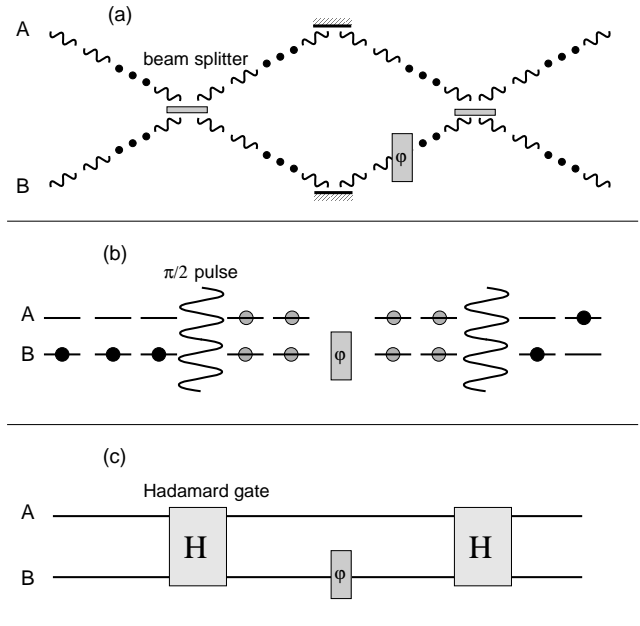} 
\end{center}
\caption{
Three distinct
representations of phase measurement:
(a) a Mach-Zehnder interferometer,
(b) Ramsey spectroscopy,
and (c) a generic quantum logic gate.
The two basis states of a qubit,
$|0\ra$ and $|1\ra$, may be regarded as the atomic
two levels, or the two paths in a Mach-Zehnder interferometer.
The state $|\varphi\rangle$ can be regarded as
a single photon state just before the second beam splitter 
in the Mach-Zehnder interferometer, or the single atom state
just before the second $\pi/2$-pulse in the Ramsey interferometer.
}
\end{figure}
In all three protocols, the initial state is transformed
by a discrete Fourier transform (beam splitter, $\pi/2$-pulse or
Hadamard), then picks up a relative phase, and is transformed back
again. The Hadamard transform is the standard (two-dimensional)
``quantum'' finite Fourier transform, such as used in the
implementation of Shor's algorithm\cite{ekert96}. The phase shift,
which is hard to measure directly, is applied to the transformed
basis. The result is a bit flip in the initial basis $\{
|0\rangle,|1\rangle\}$, and this is readily measured. We call the
formal equivalence between these three systems the 
{\em quantum Rosetta stone}\cite{lee02}. 
In discussing quantum computer circuits
with researchers from the fields of quantum optics or atomic 
clocks, we find the ``quantum Rosetta stone'' a useful tool. For
example, in a Ramsey-type atom interferometer, noting that ${\hat
  A}\equiv \sigma_x = H \sigma_z H$, and
\be
\hat{A}_N = {\mbox{\Large $\otimes$}}_{k=1}^N  \hat{A}^{(k)}
=
\left( {\mbox{\Large $\otimes$}}_{k=1}^N  H^{(k)}\right)
\hat{A}_N^\prime 
\left({\mbox{\Large $\otimes$}}_{k=1}^N  H^{(k)}\right),
\ee
we need to measure $\hat{A}_N^\prime = {\mbox{\Large
    $\otimes$}}_{k=1}^N  \sigma_z^{(k)}$ after the second beam
splitter (see Fig.~2). On the other hand, a direct optical measurement
of $\hat{A}_N$ corresponds to an $N$-photion absorption scheme in
quantum lithography.

\begin{figure}[t]
\epsfxsize=18pc
\begin{center}
\epsfbox{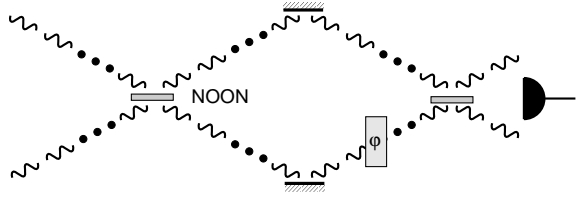}
\end{center}
\caption{
Quantum interferometry. 
{\tt NOON} represents the maximally entangled state
$|N,0\ra +|0,N\ra$ as given in Eq.~(\ref{noon}).
After the second beam splitter, an obsevable ${\hat A}_N^\prime$
is measured.}
\end{figure}

\begin{figure}[t]
\epsfxsize=14pc
\begin{center}
\epsfbox{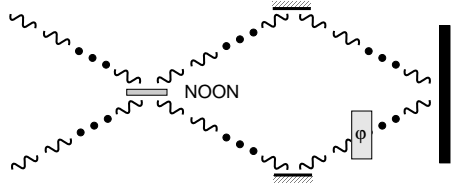}
\end{center}
\caption{
Quantum lithography. 
{\tt NOON} represents the maximally path-entangled state.
$N$-photon absorption scheme is analogous to a
direct measurement of the observable ${\hat A}_N$.
}
\end{figure}

Such an enhancement of a factor of $\sqrt{N}$ in quantum metrology is
known to be the best possible precision permitted by the uncertainty
principle\cite{ou96}. It is also interesting to see that exploiting
quantum correlations yields a $\sqrt{N}$ enhancement in Grover's
search algorithm, which has also been shown to be optimal\cite{zalka99}. 
Is this $\sqrt{N}$ enhancement over the classical protocols universal?
It certainly does seem so. But then Shor's algorithm shows its
exponential improvement over the best {\em known} classical
algorithm. 

We like to end this article with the ``Williams-Dowling Inverse-Shor
Conjecture''\cite{williams00}. Assume that the best {\em known}
classical factoring algorithm is the best {\em possible} one. Then,
according to Conjecture 1, the quantum Rosetta stone tells us that
there must exist an interferometric measurement strategy for phase
sensitivity that is exponentially bettter than the shot-noise limit
(the best classical strategy). However, we know that this is false.
Conjecture 2, therefore, tells us that according to the quantum
Rosetta stone, Shor's algorithm is a $\sqrt{N}$ improvement over the
{\em best, but unknown} classical protocol. Thus, 
there exists a
classical algorithm that is exponetially faster than the {best known}
one,
though quadratically slower than the quantum algorithm!

\section*{Acknowledgments}
This work was carried out by the Jet Propulsion Laboratory,
California Institute of Technology, 
under a contract with the National Aeronautics
and Space Administration.
We wish to thank
C.\ P.\ Williams, 
and D.\ J.\ Wineland for stimulating discussions. 
We would like to acknowledge support from the ONR,
ARDA, NSA, and DARPA.
P.K.\ and H.L.\ would also like to acknowledge
the National Research Council.


\end{document}